\documentclass[10pt]{article}
\newcommand{\beq}{\begin{equation}}
\newcommand{\eeq}{\end{equation}}
\newcommand{\be}{\begin{equation}}
\newcommand{\ee}{\end{equation}}
\newcommand{\beqa}{\begin{eqnarray}}
\newcommand{\eeqa}{\end{eqnarray}}
\newcommand{\beqar}{\begin{eqnarray*}}
\newcommand{\eeqar}{\end{eqnarray*}}
\newcommand{\bea}{\begin{eqnarray}}
\newcommand{\eea}{\end{eqnarray}}

\usepackage{amsmath}
\usepackage{amsmath, dsfont}
\usepackage{amssymb}
\usepackage[dvips]{graphics}
\usepackage[dvipsnames]{xcolor}
\usepackage{epsfig}
\usepackage{calc}
\usepackage{amsfonts}
\usepackage{graphicx}
\usepackage[ansinew]{inputenc}

\title{Rotating stealth black holes with a cohomogeneity$-1$ metric}
\author{Olaf Baake$^{\dag\,\ddagger}$\footnote{olaf.baake-at-inst-mat.utalca.cl} and Mokhtar Hassaine$^{\dag}$\footnote{hassaine-at-inst-mat.utalca.cl}\\
\normalsize{$^{\dag}$Instituto de
Matem\'atica (INSTMAT), Universidad de Talca, Casilla 747, Talca 3460000,
Chile} \\ \normalsize{$^{\ddagger}$Centro de Estudios Cientificos (CECs), Av.
Arturo Part 514, Valdivia 5090000, Chile.} \\\\
} \textheight 16 cm\textwidth 16 cm
\let\ssection=\section
\renewcommand{\section}{\setcounter{equation}{0}\ssection}

\begin{document}
\maketitle
\begin{abstract}
In five dimensions we consider a general shift symmetric and
parity preserving scalar tensor action that contains up to second order covariant
derivatives of the scalar field. A rotating stealth black hole solution
is constructed where the metric is given by the
Myers-Perry spacetime with equal momenta and the scalar field is
identified with the Hamilton-Jacobi potential. This nontrivial scalar field has an extra hair associated with the
rest mass of the test particle, and the solution does not require any fine tuning of the coupling functions of
the theory. Interestingly enough, we show that the disformal transformation, generated by this scalar field, and with
a constant degree of disformality, leaves invariant (up to diffeomorphisms) the Myers-Perry metric with equal momenta.
This means that the hair of the scalar field, along with the constant disformality parameter, can be consistently absorbed into
further redefinitions of the mass and of the single angular parameter of the disformed metric.
These results are extended in higher odd dimensions with a Myers-Perry metric for which all the momenta are equal. The key of the
invariance under disformal transformation of the metric is mainly the cohomogeneity$-1$ character of the Myers-Perry metric with
equal momenta. Starting from this observation, we consider a general class of cohomogeneity$-1$ metrics in arbitrary dimension,
and we list the conditions ensuring that this
class of metrics remain invariant (up to diffeomorphisms) under a disformal
transformation with a constant degree of disformality and with a scalar field
with constant kinetic term. The extension to the Kerr-de Sitter case is also considered where
it is shown that rotating stealth solutions may exist provided some fine tuning of the coupling functions of the scalar tensor theory.
\end{abstract}

%%%%%%%%%%%%%%%%%%%%%%%%%%%%%%%%%%
\section{Introduction}
%%%%%%%%%%%%%%%%%%%%%%%%%%%%%%%%%%
Even though the detection of gravitational waves \cite{Abbott:2016blz}
has raised Einstein's four-dimensional General Relativity (GR)
to an exceptional position, this should not slow down our desire of exploring
the theories of gravity in higher dimensions, as well as to study its black
hole solutions. The interests in these studies are numerous and diverse.
For example one can mention the gauge/gravity duality \cite{Maldacena:1997re} which allows to relate the properties of black holes in some dimension to the properties of strongly coupled quantum field theories defined in some lower dimension. For
example, by applying the AdS/CFT machinery with a five-dimensional AdS black hole, the authors of Ref. \cite{Kovtun:2004de} were able to show that the ratio of the shear viscosity and the volume density of entropy was close to a certain universal constant, which was further confirmed at the Relativistic Heavy Ion Collider. From a different
point of view, it is undeniable that, using a purely mathematical approach, the study of higher-dimensional black holes has required the development of new mathematical tools with significant benefit for the scientific community. For a nice review on higher-dimensional
black holes see Ref. \cite{Emparan:2008eg}.
In the same spirit we deem it important to accompany the studies of higher-dimensional black holes with modifying GR in order to explore new promising theoretical possibilities in the realm of gravity. From this angle, scalar tensor theories have attracted a great deal of attention over the past two decades. These theories can be considered as one of the simplest modifications to the theory of gravity, since one only needs to introduce a single scalar field in addition to the metric. One of the pioneering works in this context was provided by Horndeski in the seventies where he presented the
most general scalar-tensor theory with second order equations of motion in four dimensions \cite{Horndeski:1974wa}. This requirement of not having more than two derivatives is a sufficient condition that prevents the theory to have a Hamiltonian that is
unbounded from below. More recently, it was shown that this virtue of the Horndeski theories can also be extended to scalar tensor theories which have higher than second order equations of motion. These latter are now known as the Degenerate Higher Order Scalar Tensor Theories (DHOST theories) \cite{Langlois:2015cwa, Crisostomi:2016tcp, Motohashi:2016ftl}. Many interesting solutions have been found concerning these DHOST theories, see e.g. \cite{Babichev:2017guv, Babichev:2017lmw, Charmousis:2019vnf, Babichev:2016kdt, Kobayashi:2019hrl, Babichev:2020qpr}. While the above-mentioned relevant features are inherent in the four-dimensional DHOST theories, we find it interesting to also consider the extension of general scalar tensor theories in higher dimensions in order to explore, among other things, their possible rotating black hole solutions.

This is precisely the aim of the present paper where we will be investigating the possibility of constructing rotating black hole solutions in these modified gravity theories. As is known, this task is highly non-trivial, especially for scalar-tensor theories, where in general the complexity and the high degree of non-linearity of the field equations almost make it impossible to find spinning solutions. Starting from
this observation, we will approach the problem from a very particular
perspective, which can be summarized as follows. We will start by fixing the metric background to be a vacuum rotating black hole spacetime, and we will investigate whether this spacetime can be endowed with a non-trivial scalar field such that the
full equations of motion are satisfied. Such solutions may be identified with the so-called stealth solutions, see Refs. \cite{AyonBeato:2004ig, AyonBeato:2005tu} for the original works on stealth configurations. In the present case, we will show that the vacuum Myers-Perry metric \cite{Myers:1986un} in higher odd dimensions, with equal
angular momenta, can accommodate a non-trivial scalar field in such a way that the resulting scalar tensor configuration will satisfy the complete field equations of some general scalar tensor theories containing, in particular, the DHOST branch, and also
including the sector with unitary speed of gravitational waves. Regarding the scalar field solution, we will see that its expression merges with that of the Hamilton-Jacobi action
\cite{Frolov:2002xf} in which the azimuthal conserved quantities are zero, and the energy is equal to the test particle mass.
More precisely, the scalar field is shown to be linearly time-dependent with a radial dependence in such way that its kinetic term is a constant given by minus the square of the test particle mass. These conditions are similar to those found for the disformed Kerr metric in \cite{Charmousis:2019vnf, Anson:2020trg, BenAchour:2020fgy}, where
it was shown that the restriction on the energy ensures the scalar hair to be well defined from the event horizon up to asymptotic infinity, while the vanishing of the azimuthal conserved quantities guarantees the regularity at the poles. Note that such an ansatz for the scalar field has been proven to be fruitful for finding solutions of Horndeski/DHOST scalar tensor theories, see Ref. \cite{Babichev:2013cya} and also \cite{Kobayashi:2014eva, Bravo-Gaete:2013dca, Minamitsuji:2019shy}.

Following Refs. \cite{Anson:2020trg, BenAchour:2020fgy}, we will make use
of the stealth scalar field solution for constructing disformal versions of the metric solution. Surprisingly, we will establish that the odd-dimensional Myers-Perry spacetimes with equal momenta remain invariant (up to diffeomorphisms and redefinitions
of the constants) under a disformal transformation generated by the stealth scalar field
with a constant degree of disformality. This result is in itself intriguing since the disformal transformations are supposed to map solutions of some classes of scalar tensor theories to other (different) classes, and the resulting disformed metrics
are in general quite different from the original ones, \cite{Zumalacarregui:2013pma,Babichev:2017lmw}. To illustrate this assertion, we mention the works of Refs. \cite{Anson:2020trg, BenAchour:2020fgy} where a disformal version of the Kerr spacetime with a regular scalar field was constructed, and where the disformed Kerr metric turned out to be neither Ricci flat nor circular \cite{Anson:2020trg}. Note that disformal transformations in the case of static metrics were previously considered in \cite{BenAchour:2019fdf}.

It is clear that the possibility of constructing rotating stealth black hole configurations, and that the corresponding scalar field leaving the metric invariant under a constant disformal transformation, are highly correlated with the particular symmetries of the metric. Indeed, it is well-known that the Myers-Perry line element in $D$ dimensions can be shown to have an isometry group identified with $\mathbb{R}\times U(1)^n$ where $\mathbb{R}$ corresponds to time translations and $n = [(D - 1)/2]$. Nevertheless, this symmetry group in odd dimension, $D=2N+3$, with all its angular momenta equal, $a_i=a$, is extended to a bigger symmetry group given by $\mathbb{R}\times U(N +1)$. As a direct consequence of this symmetry enhancement, the odd-dimensional Myers-Perry spacetime with equal momenta is cohomogeneity$-1$, which is to say that it only non-trivially depends on
a single coordinate. We may note that the cohomogeneity$-1$ character of the Myers-Perry metrics has been proven to be of great importance in order to study the stability of theses particular odd-dimensional Myers-Perry spacetimes, see \cite{Dias:2010eu}, where it was concluded that there was no evidence of instability in five and seven dimensions in contrast with the nine-dimensional case where an instability was found. In the present work, one can claim with sincerity that the cohomogeneity$-1$ property of the Myers-Perry metrics with equal angular momenta is clearly responsible for its disformal invariance.
We will go further in this direction by considering a general class of cohomogeneity$-1$ metrics in arbitrary dimension, and by establishing a list of conditions for the metric, ensuring its invariance (up to diffeomorphisms) under a disformal transformation with a constant degree of disformality, and with a scalar field whose kinetic term is constant.

The extension to the Kerr-de Sitter case is also considered where
it is shown that rotating stealth solutions may exist provided some fine tuning of the coupling functions of the scalar tensor theory.

The plan of the paper is organized as follows. In Sec. $2$, we will introduce the scalar tensor theories under consideration and present their field equations in the particular case where the scalar field is taken to have a constant kinetic term. We will establish that for a vacuum black hole metric the complete set of field equations can be satisfied,  provided that the scalar field solves a unique constraint equation which, as we will see, can be solved in odd dimensions. In Sec. $3$, we will start by considering the five-dimensional case where the rotating stealth black hole will be constructed on the Myers-Perry spacetime with equal angular momenta. We will also see that the scalar field leaves the metric invariant under a constant disformal transformation. These results will then be generalized in higher odd dimensions where the Myers-Perry spacetime
with equal momenta is cohomogeneity$-1$. In Sec. $4$, starting from this observation, we consider a general class of
cohomogeneity$-1$ metrics in arbitrary dimension, and we list the conditions ensuring this
class of metrics to remain invariant (up to diffeomorphisms) under a constant disformal
transformation. The last section is concerned with our conclusions while an Appendix is devoted to extend our results in presence of a cosmological constant.

%%%%%%%%%%%%%%%%%%%%%%%%%%%%%%
\section{Set up of the theory}
%%%%%%%%%%%%%%%%%%%%%%%%%%%%%%
In the present work, we will be concerned with the following shift symmetric, and parity preserving scalar tensor action that contains up to second order covariant derivatives of the scalar field
\begin{eqnarray}
S[g,\phi]=\int
d^Dx\sqrt{-g}&\Big[G(X)R+A_1(X)\left[\phi_{\mu\nu}\phi^{\mu\nu}-(\Box\phi)^2\right]+A_3(X)\Box\phi\,\phi^{\mu}\phi_{\mu\nu}\phi^{\nu}
\nonumber\\
&+A_4(X)\phi^{\mu}\phi_{\mu\nu}\phi^{\nu\rho}\phi_{\rho}+A_5(X)\left(\phi^{\mu}\phi_{\mu\nu}\phi^{\nu}\right)^2\Big],
\label{actionSST}
\end{eqnarray}
Here the coupling functions $G, A_1, A_3, A_4$ and $A_5$ are functions of the kinetic term, $X=g^{\mu\nu}\phi_{\mu}\phi_{\nu}$, in order
to ensure the shift symmetry $\phi\to\phi+\mbox{cst}$. Also, for simplicity we have defined $\phi_{\mu}=\nabla_{\mu}\phi$ and
$\phi_{\mu\nu}=\nabla_{\mu}\nabla_{\nu}\phi$. As mentioned in the introduction, the action (\ref{actionSST}) can propagate healthy degrees of freedom in four dimensions, provided the functions $A_4$ and $A_5$ are constrained by
\begin{eqnarray*}
\label{a4a5} A_4&=&\frac{1}{8(G-XA_1)^2} \left\{ 4G \left[
3(-A_1+2G')^2-2A_3G\right] -A_3X^2(16A_1G'+A_3 G) \right.
\nonumber\\
&& \left. + 4X \left[
3A_1A_3G+16A_1^2G'-16A_1(G')^2-4A_1^3+2A_3G G' \right]
\right\},
\nonumber\\
A_5&=&\frac{1}{8(G-X A_1)^2} (2A_1-XA_3-4G')
\left(A_1(2A_1+3XA_3-4G')-4A_3G\right),
\end{eqnarray*}
but for our specific task, we will not {\it a priori} consider such restrictions on the coupling functions.

Neither will we write down the equations of motion in all of their generality, but instead restrict their expressions in the case of a constant kinetic term $X=\mbox{cst}$. In doing so, the equations arising from the variation of the
general scalar tensor field action (\ref{actionSST}) with respect to the metric reduce to
\begin{eqnarray}
&&G(X)G_{\mu\nu}+G^{\prime}(X) R\phi_{\mu}\phi_{\nu}-\frac{1}{2}A_3(X)\Big[(\Box\phi)^2-(\phi_{\alpha\beta})(\phi^{\alpha\beta})\Big]\phi_{\mu}\phi_{\nu}
+\frac{1}{2}A_3(X)\Big[R_{\alpha\beta}\phi^{\alpha}\phi^{\beta}\Big]\phi_{\mu}\phi_{\nu}\nonumber\\
&&+A_1(X)\Bigg[-R_{\nu\lambda}\phi_{\mu}\phi^{\lambda}-R_{\mu\lambda}\phi_{\nu}\phi^{\lambda}-
\frac{1}{2}g_{\mu\nu}\left[(\Box\phi)^2-(\phi_{\alpha\beta})(\phi^{\alpha\beta})\right]+g_{\mu\nu}\left[R_{\lambda\rho}\phi^{\lambda}\phi^{\rho}\right]
+\phi_{\mu\nu}\Box\phi+ \phi^{\lambda}\phi_{\lambda\mu\nu}\Bigg]\nonumber\\
&&-A'_{1}(X)\left[(\Box\phi)^2-(\phi_{\alpha\beta})(\phi^{\alpha\beta})\right]\phi_{\mu}\phi_{\nu}=0,
\label{feqs}
\end{eqnarray}
while its variation with respect to the scalar field is a conserved current equation given by $\nabla_{\mu}J^{\mu}=0$, with
{\small
\begin{eqnarray}
J^\mu = 2 \left( G'(X) R - \left[ A_1'(X) + \frac{1}{2} A_3(X) \right] \Big[(\Box\phi)^2-(\phi_{\alpha\beta})(\phi^{\alpha\beta})\Big] + \frac{1}{2}A_3(X)\Big[R_{\alpha\beta}\phi^{\alpha}\phi^{\beta}\Big] \right)\phi^\mu - 2 A_1(X) R^{\mu\nu}\phi_\nu.
\label{Jeqs}
\end{eqnarray}}
Note that this conservation equation is a direct consequence of the shift symmetry of the action.

Further, it is remarkable to note that for any any vacuum metric solution $R_{\mu\nu}=0$, the field equations (\ref{feqs}-\ref{Jeqs}) will be automatically fulfilled, provided that the scalar field satisfies (in addition of having a constant kinetic term) the following two conditions
\begin{subequations}
\label{cccc}
\begin{eqnarray}
\label{cond1z}
&&(\Box\phi)^2-\left(\phi_{\mu\nu}\right)\left(\phi^{\mu\nu}\right)=0,\\
&&\phi_{\mu\nu}\,\Box\phi+ \phi^{\lambda}\phi_{\lambda\mu\nu}=0,
\label{cond2z}
\end{eqnarray}
\end{subequations}
and this without imposing any restrictions on the coupling functions, $G(X), A_1(X), A_3(X), A_4(X)$ and $A_5(X)$. On the other hand, since the kinetic term of the scalar field is constant, $\phi_{\mu}\phi^{\mu}=\mbox{cst}$, it is easy to see that the trace of the second equation
(\ref{cond2z}) yields the first condition (\ref{cond1z}). As a direct consequence, for a vacuum metric it will be sufficient for the scalar field to have a constant kinetic term and to satisfy the following tensorial equation
\begin{eqnarray}
\phi_{\mu\nu}\,\Box\phi+ \phi^{\lambda}\phi_{\lambda\mu\nu}=0,
\label{cond}
\end{eqnarray}
to ensure the field equations (\ref{feqs}-\ref{Jeqs}) to be satisfied for any coupling functions. This observation will be our guiding principle in order to construct stealth
rotating black hole solutions of the general scalar tensor theory defined by (\ref{actionSST}). In presence of matter sources, conditions ensuring the construction of solutions in general quadratic higher order scalar-tensor theories with and without cosmological constant have been derived in \cite{Takahashi:2020hso}. However, in contrast to the present case, these conditions put constraints on the coupling functions of the theories instead of the scalar field (for example requiring that $A_1$ and $A_2$ vanish at the constant value of $X$). \\
Now, in order to construct the corresponding stealth scalar field, $\phi$, we must first make sure that its kinetic term is constant. A simple option would be to identify the scalar field with the Hamilton-Jacobi potential, $S$, as is done in \cite{Charmousis:2019vnf}, i.e.
\begin{eqnarray}
\phi\equiv{\cal S},
\label{sfHJ}
\end{eqnarray}
where ${\cal S}$ satisfies the Hamilton-Jacobi equation of a free particle of mass $m$,
\begin{eqnarray}
g^{\mu\nu}\,\partial_{\mu}{\cal S}\, \partial_{\nu}{\cal S}=-m^2.
\label{HJeqssss}
\end{eqnarray}
This hypothesis on the scalar field (\ref{sfHJ}) is also useful to take advantage on the known results on the integrability of
the Hamilton-Jacobi equations. It is also clear that, in this representation,
the constant value of the kinetic term would be given by minus the
square of the mass of the particle, i.e. $\phi_{\mu}\phi^{\mu}=-m^2$.

In what follows, we will consider vacuum spacetime metrics (representing rotating black holes) with a scalar field identified with the corresponding Hamilton-Jacobi potential (\ref{sfHJ}), and  we will discuss under which conditions the tensorial equations (\ref{cond}) can be fulfilled.

%%%%%%%%%%%%%%%%%%%%%%%%%%%%%%%%%%%%%%%%%%%%%%%%%%%%%%%%%%%%%%%%%%%%%%%%%%%%%%%%%%%%%%
\section{Rotating stealth black holes and their disformal transformations }
%%%%%%%%%%%%%%%%%%%%%%%%%%%%%%%%%%%%%%%%%%%%%%%%%%%%%%%%%%%%%%%%%%%%%%%%%%%%%%%%%%%%%%%%
In this section, we construct a concrete example of a rotating stealth black hole solution for the scalar theory defined by the action (\ref{actionSST}) and whose field equations for a constant kinetic term reduce to (\ref{feqs}-\ref{Jeqs}). As
shown below, this will be possible in odd dimensions, and for a Myers-Perry vacuum metric \cite{Myers:1986un} where all the angular momenta $a_i$ take a single value, $a_i=a$, together with a scalar field identified with the Hamilton-Jacobi potential (\ref{sfHJ}-\ref{HJeqssss}). Having in hand this scalar tensor solution, namely a metric, $g$, and a nontrivial scalar field, $\phi$, it will be interesting to study the disformal transformation of the metric
$$
\bar{g}_{\mu\nu}={g}_{\mu\nu}-P(\phi, X)\,\,\phi_{\mu}\,\phi_{\nu},
$$
where $P$ may be an arbitrary function of the scalar field and its kinetic term. The interest on such consideration is mainly due to the fact that such disformal transformations are known to be internal maps of the scalar tensor theories considered here (\ref{actionSST}). The special ingredient in our construction will be the fact that the the scalar field responsible for the disformed metric is related to the geodesics of Myers-Perry spacetimes (\ref{sfHJ}-\ref{HJeqssss}). In Refs. \cite{Anson:2020trg, BenAchour:2020fgy}, this construction was done for a stealth solution defined on the four-dimensional Kerr metric, and it was shown that the deviation of the disformed metric with respect to the Kerr metric is considerable. Indeed, the disformed Kerr metric was shown to be neither Ricci flat, nor circular, and obviously no longer a vacuum metric. In the present case, we will see that the disformal transformation of the Myers-Perry metric with equal angular momenta, denoted by ${g}_{\mu\nu}^{\mbox{{\tiny MP,}} a_i=a}$, and with a constant disformality parameter, $P$, i.e.
\begin{eqnarray}
\bar{g}_{\mu\nu}={g}_{\mu\nu}^{\mbox{{\tiny MP,}} a_i=a}-P\,\phi_{\mu}\,\phi_{\nu},
\label{disftr}
\end{eqnarray}
is diffeomorphic to itself, that is $\bar{g}_{\mu\nu}\underset{\mbox{\tiny{diffeo}}}{\sim} {g}_{\mu\nu}^{\mbox{{\tiny MP,}} a_i=a}$.

We first analyze in detail the five-dimensional case before considering the
extension to higher odd dimensions.

%%%%%%%%%%%%%%%%%%%%%%%%%%%%%%%%%%%%%%%%%%%%%%%%%%%%%%%%%%%%%%%%%%%%%%%%%%%%%%%%%%%%%%%%%%%%%%%%%%%%%
\subsection{Stealth on the five-dimensional Myers-Perry spacetime and its (invariant) disformed transformation}
%%%%%%%%%%%%%%%%%%%%%%%%%%%%%%%%%%%%%%%%%%%%%%%%%%%%%%%%%%%%%%%%%%%%%%%%%%%%%%%%%%%%%%%%%%%%%%%%%%%
As we have seen in the previous section, any vacuum metric together with a scalar field satisfying the tensorial equation (\ref{cond}) will be a solution of the full field equations (\ref{feqs}-\ref{Jeqs}) without any conditions on the coupling functions. In four dimensions, this problem was considered in Ref. \cite{Charmousis:2019vnf}, where the
authors constructed a stealth solution defined on the Kerr(-de Sitter) metric. Nevertheless, in this case we would like to underline that, as the stealth scalar field does not fulfill the conditions (\ref{cond}), restrictions on the coupling functions are necessary. We now turn to the five-dimensional situation where we will notice that this kind of restrictions can be circumvented thanks to the symmetries of the vacuum metric.

The five-dimensional Myers-Perry solution \cite{Myers:1986un} of the vacuum Einstein equations, $R_{\mu\nu}=0$, in Boyer-Lindquist coordinates, $(t,r,\theta,\varphi,\psi)$, reads
\begin{eqnarray}\label{5mpbh}
ds^2_{\mbox{{\tiny MP}}} &=& -\left(1-\frac{2M}{\rho^2} \right)dt^2 + \frac{r^2
\rho^2}{\Delta}dr^2 +\rho^2 d \theta^2 +\frac{4 a M
\sin^2\theta}{\rho^2} dt d\varphi  + \frac{4 b M
\cos^2\theta}{\rho^2} dt d\psi \nonumber\\ && + \frac{4 a b M \sin^2
\theta \cos^2 \theta}{\rho^2} d\varphi d\psi  + \sin^2\theta
\left(r^2+a^2+\frac{2M a^2 \sin^2 \theta}{\rho^2} \right)d\varphi^2
\nonumber\\ && + \cos^2\theta \left(r^2+b^2+\frac{2M b^2
\cos^2\theta}{\rho^2} \right)d\psi^2,
\end{eqnarray}
where we have defined
\begin{eqnarray}\label{5delta}
\Delta = (r^2+a^2)(r^2+b^2)-2M r^2,  \quad \quad   \rho^2 = r^2+a^2
\cos^2\theta + b^2 \sin^2\theta.
\end{eqnarray}
The metric~(\ref{5mpbh}) is characterized by its mass, $M$, and two
angular momenta denoted by $a$ and $b$. Once the vacuum metric is fixed,
following the strategy outlined in the previous section, we now identify the scalar field with the Hamilton-Jacobi potential associated to the five-dimensional Myers-Perry solution. Fortunately, it was shown in Ref. \cite{Frolov:2002xf} that the Hamilton-Jacobi equation of the Myers-Perry metric can be separated into the form
\begin{eqnarray}
{\cal S}=\frac{1}{2}m^2\lambda-Et+S_r(r)+S_{\theta}(\theta)+L_1\,\varphi+L_2\,\psi,
\label{sfHJ2}
\end{eqnarray}
where $E$ is the energy,  the $L_i$'s are the conserved quantities associated
with each rotation, and
\begin{eqnarray*}
S_r(r)=\int^r \frac{\sqrt{X(r)}}{\Delta}\,r\,dr,\qquad S_{\theta}(\theta)=\int^{\theta}\sqrt{\Theta(\theta)}\,d\theta,
\end{eqnarray*}
with {\small
\begin{eqnarray*}
&& X(r)=\Delta\Big[r^2(E^2-m^2)+(a^2-b^2)\left(\frac{L_1^2}{r^2+a^2}-\frac{L_2^2}{r^2+b^2}\right)-K\Big]+
2M(r^2+a^2)(r^2+b^2)\left[E+\frac{aL_1}{r^2+a^2}+\frac{bL_2}{r^2+b^2}\right]^2,\\
&& \Theta(\theta)=(E^2-m^2)\left(a^2\cos(\theta)^2+b^2\sin(\theta)^2\right)-\frac{L_1^2}{\sin(\theta)^2}-\frac{L_2^2}{\cos(\theta)^2}+K.
\end{eqnarray*}}
Here $K$ is a constant allowing to separate the equations of motion, and which plays a role
analogous to Carter's constant, \cite{Carter:1968rr}.

For the identification of the scalar field with the Hamilton-Jacobi potential (\ref{sfHJ}-\ref{sfHJ2}), one can show after some calculations (which are somewhat too cumbersome
to report them here) that the tensorial equations (\ref{cond}) will be
satisfied if: (i) the azimuthal conserved quantities are zero, $L_1=L_2=0$, (ii) the Carter constant is zero, $K=0$, (iii) the values of the the two angular momenta of the Myers-Perry metric are equal, $b=a$, and (iv) the energy, $E$, is equal to the test particle mass, $E=m$.
In other words, the non-trivial stealth scalar field is linear in time and has a radial dependence,
\begin{eqnarray}
\phi(t,r)=-m t+S_r(r)\Longrightarrow \phi_{\mu}\phi^{\mu}=-m^2.
\label{sf}
\end{eqnarray}
These restrictions are similar to those obtained for the disformed Kerr(-de Sitter) spacetimes \cite{Charmousis:2019vnf, Anson:2020trg, BenAchour:2020fgy} in order to deal with a regular scalar field. We then conclude that a rotating stealth black hole solution of the scalar tensor theory (\ref{actionSST}) with arbitrary coupling functions, $G(X), A_1(X), A_3(X), A_4(X)$ and $A_5(X)$, can be given by
\begin{eqnarray}\label{5mpbha=b}
ds^2_{\mbox{{\tiny MP}}, b=a} &=& -\left(1-\frac{2M}{r^2+a^2}
\right)dt^2 + \frac{r^2 (r^2+a^2)}{(r^2+a^2)^2-2Mr^2} dr^2+(r^2+a^2)
d \theta^2 +\frac{4 a M \sin^2\theta}{r^2+a^2} dt
d\varphi  \nonumber\\
&&+\frac{4 a M \cos^2\theta}{r^2+a^2} dt d\psi  + \frac{4 a^2 M \sin^2 \theta \cos^2 \theta}{r^2+a^2} d\varphi
d\psi + \sin^2\theta \left(r^2+a^2+\frac{2M a^2 \sin^2
\theta}{r^2+a^2} \right)d\varphi^2 \nonumber\\ && + \cos^2\theta
\left(r^2+a^2+\frac{2M a^2 \cos^2\theta}{r^2+a^2} \right)d\psi^2,
\end{eqnarray}
together with a scalar field defined by
\begin{eqnarray}
\phi(t,r)=-m t-\sqrt{2M}m\int \frac{r(r^2+a^2)}{(r^2+a^2)^2-2Mr^2}dr.
\label{sf5dMP}
\end{eqnarray}
Many comments can be made concerning this solution. Firstly, it is important to stress the importance of the linear time dependence of the scalar field (\ref{sf5dMP}) which ensures  the existence of a non-trivial stealth configuration. Indeed, eliminating the time dependency of the scalar field would amount to considering an identically-to-zero scalar field. Leaving aside the physical interpretation of the constant $m$ in (\ref{sf5dMP}), it can be set to one by exploiting the fact that the tensorial equation (\ref{cond}) is quadratic in the scalar field, and hence invariant under a rescaling by a constant, i.e. $\phi\to \frac{1}{m}\phi$.
Also, the Myers-Perry metric with equal momenta enjoys an extension of its symmetry given by $U(1)\times SU(2)=U(2)$, which gives rise to its cohomogeneity$-1$ characteristic. Finally, we shall mention that in the non-rotating limit, $a=0$, the solution reduces to a stealth configuration defined on the Schwarzschild spacetime.

Surprisingly, the non-trivial scalar field (\ref{sf5dMP}) will be shown to leave the cohomogeneity$-1$ metric (\ref{5mpbha=b}) invariant (up to diffeomorphisms) through a disformal transformation (\ref{disftr}) with a constant degree of disformality $P$. Indeed, in order to show this result explicitly, the disformed Myers-Perry metric with equal momenta, as defined in (\ref{disftr}), becomes
\begin{eqnarray}
\label{disf1}
&&d{\bar{s}}_{{\tiny\mbox{disf. MP}}}^2= -\left(1-\frac{2M}{r^2+a^2}+Pm^2 \right)dT^2 +
\frac{r^2 (r^2+a^2)}{(r^2+a^2)^2-\frac{2M}{Pm^2+1}r^2+\frac{2MPa^2
m^2}{Pm^2+1}} dr^2+(r^2+a^2) d \theta^2 \nonumber\\
&&+\frac{4 a M \sin^2\theta}{r^2+a^2} dT d\Phi  +\frac{4 a M
\cos^2\theta}{r^2+a^2} dT d\Psi+ \frac{4 a^2 M \sin^2 \theta \cos^2
\theta}{r^2+a^2} d\Phi d\Psi  + \sin^2\theta
\left(r^2+a^2+\frac{2M a^2 \sin^2 \theta}{r^2+a^2} \right)d\Phi^2
\nonumber\\ && + \cos^2\theta \left(r^2+a^2+\frac{2M a^2
\cos^2\theta}{r^2+a^2} \right)d\Psi^2,
\end{eqnarray}
where, in order to eliminate the undesirable cross-terms, we have introduced the new variables $T, \Phi$ and $\Psi$ by means of
\begin{eqnarray*}
dt=dT+k_0(r)dr,\qquad d\varphi=d\Phi+k_1(r)dr,\qquad d\psi=d\Psi+k_1(r)dr,
\end{eqnarray*}
with
\begin{eqnarray*}
&& k_0(r)=\frac{Pm^2\sqrt{2M}r(r^2+a^2)\left[(r^2+a^2)^2+2Ma^2\right]}{\left[(r^2+a^2)^2-2Mr^2\right]\left[(r^2+a^2)^2(Pm^2+1)-2M(r^2-Pm^2 a^2)\right]},\\
&& k_1(r)=\frac{-(2M)^{3/2}Pm^2r^2(r^2+a^2)}{\left[(r^2+a^2)^2-2Mr^2\right]\left[(r^2+a^2)^2(Pm^2+1)-2M(r^2-Pm^2 a^2)\right]}.
\end{eqnarray*}
Finally, under the following redefinitions
\begin{eqnarray}
\bar{t}=T\sqrt{1+Pm^2},\qquad \bar{r}=\sqrt{r^2-Pm^2a^2},\qquad \bar{M}=\frac{M}{1+Pm^2},\qquad \bar{a}^2=\frac{a^2}{1+Pm^2},
\label{redfs}
\end{eqnarray}
it is easy to see that the disformed metric (\ref{disf1}) is nothing but the Myers-Perry metric (\ref{5mpbha=b}) with equal
angular momenta $\bar{a}$ and with mass $\bar{M}$. This result is surprising by itself, since one would expect that the hair of the scalar field, $m$, and the constant disformality factor, $P$, would have a certain impact in the disformed metric, but this is not the case, since both parameters can be consistently absorbed into the redefinitions of the coordinates, and into the physical constants (\ref{redfs}). This is clearly in contrast with the four-dimensional disformed Kerr metric \cite{Anson:2020trg}, where the deviations from General Relativity are strongly codified by the disformality coefficient $P$, which cannot be absorbed.

%In summary, we conclude that the five-dimensional Myers-Perry solution \cite{Myers:1986un}, with equal angular momenta,
%remains invariant, up to diffeomorphism and some redefinitions of the integration constants (\ref{redfs}), under the
%disformal transformation (\ref{disftr}) with a constant degree of disformality and with a scalar field given by (\ref{sf5dMP}).

Before extending these results to higher odd-dimensions, in which the Myers-Perry spacetime can be as well a cohomogeneity$-1$ metric, we would like
to propose a geometrical explanation of the disformal invariance. Generically, for a disformal transformation of the form
$$
\bar{g}_{\mu\nu}=g_{\mu\nu}-P\phi_{\mu}\,\phi_{\nu},
$$
with $P$ being constant, and with a scalar field $\phi$ such that $X=g^{\mu\nu}\partial_{\mu}\phi\partial_{\nu}\phi=\mbox{cst}$, the Riemann and Ricci
tensors of the disformed metric $\bar{g}_{\mu\nu}$ can be expressed as
\begin{eqnarray}
\bar{R}^{\alpha}_{\,\,\beta\mu\nu}={R}^{\alpha}_{\,\,\beta\mu\nu}+2\nabla_{[\mu} {\cal K}^{\alpha}_{\,\,\nu]\beta}+
2{\cal K}^{\alpha}_{\,\,\gamma[\mu}{\cal K}^{\gamma}_{\,\,\nu]\beta}\Longrightarrow \bar{R}_{\mu\nu}={R}_{\mu\nu}+2\nabla_{[\alpha} {\cal K}^{\alpha}_{\,\,\mu]\nu}+
2{\cal K}^{\alpha}_{\,\,\gamma[\alpha}{\cal K}^{\gamma}_{\,\,\mu]\nu},
\label{ricci}
\end{eqnarray}
where we have defined
\begin{eqnarray}
{\cal K}^{\alpha}_{\,\,\mu\nu}:=\bar{\Gamma}^{\alpha}_{\mu\nu}-{\Gamma}^{\alpha}_{\mu\nu}=
\bar{g}^{\alpha\lambda}\Big(\nabla_{(\mu}\bar{g}_{\nu)\lambda}-\frac{1}{2}\nabla_{\lambda}\bar{g}_{\mu\nu}\Big).
\end{eqnarray}
We know that for a constant degree of disformality, $P$, the expression of the tensor ${\cal K}^{\alpha}_{\,\,\mu\nu}$ reduces to
\begin{eqnarray}
{\cal K}^{\alpha}_{\,\,\mu\nu}=-\frac{P}{1-PX}\phi^{\alpha}\phi_{\mu\nu}.
\end{eqnarray}
On the other hand, for $X=\mbox{cst}$, one can easily establish that each term of ${\cal K}^{\alpha}_{\,\,\gamma[\alpha}{\cal K}^{\gamma}_{\,\,\mu]\nu}$ vanishes, as well as
$\nabla_{\mu} {\cal K}^{\alpha}_{\,\,\alpha\nu}$, and hence the expression of the Ricci tensor of the disformed metric (\ref{ricci}) reduces to
\begin{eqnarray}
\bar{R}_{\mu\nu}={R}_{\mu\nu}+\nabla_{\alpha} {\cal K}^{\alpha}_{\,\,\mu\nu}={R}_{\mu\nu}-\frac{P}{1-PX}\nabla_{\alpha}
\Big(\phi^{\alpha}\phi_{\mu\nu}\Big).
\label{riccibar}
\end{eqnarray}
This last expression is noting but the Ricci tensor of any disformed metric constructed by means of a constant degree of disformality, $P$, and with a scalar field whose kinetic term, $X$, is constant. In our particular situation, since the Myers-Perry metric is a vacuum metric solution, the expression (\ref{riccibar}) becomes
\begin{eqnarray}
\bar{R}_{\mu\nu}=-\frac{P}{1-PX}\nabla_{\alpha}
\Big(\phi^{\alpha}\phi_{\mu\nu}\Big)=-\frac{P}{1-PX}\left(\phi_{\mu\nu}\,\Box\phi+ \phi^{\lambda}\phi_{\lambda\mu\nu}\right),
\label{riccibar2}
\end{eqnarray}
and hence, we can conclude that the disformed metric is also a vacuum metric if and only if the following current
\begin{eqnarray}
j^{\alpha}_{\mu\nu}:=\phi^{\alpha}\phi_{\mu\nu},
\label{cur}
\end{eqnarray}
is conserved, i.e. $\nabla_{\alpha}j^{\alpha}_{\mu\nu}=0$. It is easy to see that this divergence is exactly
the tensorial equation (\ref{cond}) that ensures our construction of the stealth solution for any vacuum metric, and
for any coupling functions.  Hence, we recover that, for a scalar field satisfying the conditions (\ref{cond}) on a vacuum metric,
its disformal transformation  will be as well a vacuum metric.

%%%%%%%%%%%%%%%%%%%%%%%%%%%%%%%%%%%%%%%%%%%%%%%%%%%%%%%%%%%%%%%%%%%%%%%%%%%%%%%%%%%%%%%%%%%%%%%%%%%%%%%%%
\subsection{Stealth on the higher odd-dimensional Myers-Perry solution with equal angular momenta and its disformed transformation}
%%%%%%%%%%%%%%%%%%%%%%%%%%%%%%%%%%%%%%%%%%%%%%%%%%%%%%%%%%%%%%%%%%%%%%%%%%%%%%%%%%%%%%%%%%%%%%%%%%%%%%%%%
In five dimensions we have seen that a stealth solution defined on the Myers-Perry spacetime was possible only if the symmetry of this spacetime is enhanced to a $SU(2)$, and that this can be done by equating the two angular momenta. As a direct consequence, this Myers-Perry spacetime turns out to be a cohomogeneity$-1$ metric. This particular feature is inherent to the Myers-Perry spacetimes only in
odd dimensions, $D=2N+3$, and in the particular case of equal angular momenta, $a_i=a$. This will be our starting point, in order to generalize the previously found
rotating stealth black hole solution in higher odd dimensions ($D>5$).

In odd dimensions, $D=2N+3$, the cohomogeneity$-1$ Myers-Perry metric can be conveniently represented as
{\small
\begin{eqnarray}
\label{background}
ds^2_{\mbox{{\tiny MP}}} = -(1-f^2(r,r_M,a))dt^2 +g^2(r,r_M,a)dr^2 + h^2(r,r_M,a)[d\psi +A_j dx^j - \Omega(r,r_M,a)dt]^2 + r^2 \hat{g}_{ij} dx^i dx^j , ~~
\end{eqnarray}}
with the metric functions given by
\begin{eqnarray}
g^2(r,r_M,a) = \left(1- \frac{r_M^{2N}}{ r^{2N}} + \frac{r_M^{2N}a^2}{r^{2N+2}}\right)^{-1} &,& \quad h^2(r,r_M,a) = r^2\left( 1+ \frac{r_M^{2N}a^2}{ r^{2N+2}} \right)\,, \nonumber \\
f(r,r_M,a) =\sqrt{1- \frac{r^2}{g^2(r,r_M,a)\,h^2(r,r_M,a)}} &,& \quad \Omega(r,r_M,a) = \frac{r_M^{2N}a}{ r^{2N} h^2(r,r_M,a)}.  \nonumber
\end{eqnarray}
Here, $r_M$ is the mass radius parameter, $\hat{g}_{ij}$ is the Fubini-Study metric on $CP^{N}$ with Ricci tensor $\hat{R}_{ij} =2(N+1) \hat{g}_{ij}\,$, and $A = A_j dx^j\,$ is related to the K\"ahler form $J$ by $dA=2J$. Following closely the five-dimensional case, we look for a scalar field depending on the radial coordinate, $r$, and
linearly in time, satisfying the constraint (\ref{cond}), and with a kinetic term $X=-m^2$. Such a scalar field for the metric representation (\ref{background}) is given by
\begin{eqnarray}
\phi(t,r)=-m t+m\int \frac{g(r,r_M,a) f(r,r_M,a)}{\sqrt{1-f(r,r_M,a)^2}} dr,
\label{sfodd}
\end{eqnarray}
and hence, one can easily conclude that in odd dimensions, $D=2N+3$, the cohomogeneity$-1$ metric (\ref{background}) together with the non-trivial scalar field (\ref{sfodd}) will represent a rotating stealth solution of the field equations associated to the scalar tensor theory (\ref{actionSST}) without imposing any constraints on the coupling functions of the theory.

Continuing the analogy with the five-dimensional case, we now consider the disformed transformation of the cohomogeneity$-1$ metric (\ref{background}) using the stealth scalar field (\ref{sfodd}). In order to bring the disformed metric in the same form as (\ref{background}), we redefine the time, $t$, and the angular coordinate, $\psi$, as
\begin{eqnarray*}
&&dt\to \frac{dt}{\sqrt{1+Pm^2}}-\frac{Pm^2 f(r,r_M,a)g(r,r_M,a)}{\sqrt{1-f^2(r,r_M,a)}(1+Pm^2-f^2(r,r_M,a))}dr,\nonumber\\
&&d\psi\to d\psi-\frac{Pm^2f(r,r_M,a)g(r,r_M,a)\Omega(r,r_M,a)}{\sqrt{1-f^2(r,r_M,a)}(1+Pm^2-f^2(r,r_M,a))}dr,
\end{eqnarray*}
and, after some algebraic manipulations, the disformed Myers-Perry metric reads
\begin{eqnarray}
\label{backgrounddis}
d\bar{s}^2_{\mbox{{\tiny disf. MP}}} &=& -\Big(1-\frac{f^2(r,r_M,a)}{1+Pm^2}\Big)dt^2 +\frac{g^2(r,r_M,a) \left(1-f^2(r,r_M,a)\right) }{1-\frac{f^2(r,r_M,a)}{(1+Pm^2)}}dr^2 \nonumber\\
&&+ h^2(r,r_M,a)\Big(d\psi +A_j dx^j - \frac{\Omega(r,r_M,a)}{\sqrt{1+Pm^2}}dt\Big)^2 + r^2 \hat{g}_{ij} dx^i dx^j.
\end{eqnarray}
Now it remains to proof that this disformed metric is diffeomorphic to the original cohomogeneity$-1$ metric (\ref{background}).
It is easy to see that under the following redefinitions of the constants of integration
\begin{eqnarray}
\bar{a}=\left(1+Pm^2\right)^{\frac{1}{2}}{a},\qquad \bar{r}_M=\left(1+Pm^2\right)^{-\frac{1}{2N}}{r}_M,
\label{arm}
\end{eqnarray}
the metric functions change as
\begin{eqnarray}
\label{homogen}
&&\frac{f^2(r,r_M,a)}{1+Pm^2}=f^2(r,\bar{r}_M,\bar{a}),\quad \frac{\Omega(r,r_M,a)}{\sqrt{1+Pm^2}}=\Omega(r,\bar{r}_M,\bar{a}), \quad
h(r,r_M,a)=h(r,\bar{r}_M,\bar{a}),\nonumber\\
&&g^2(r,r_M,a)^2 \left(1-f^2(r,r_M,a)\right)=g^2(r,\bar{r}_M,\bar{a}) \left(1-f^2(r,\bar{r}_M,\bar{a})\right),
\end{eqnarray}
and hence the disformed metric (\ref{backgrounddis}) is nothing but the original Myers-Perry spacetime (\ref{background}) with the constants of integration given by $\bar{a}$ and $\bar{r}_M$, as defined in (\ref{arm}). As in the five-dimensional case,
the scalar field's hair, $m$, along with the constant disformality parameter, $P$, can be reasonably absorbed into the redefinitions of the constants of integration
(\ref{arm}). As a last remark, one can note that by redefining the constants of integration as
\begin{eqnarray}
x_M:=r_M^N,\qquad y:=\frac{1}{a},
\end{eqnarray}
the conditions (\ref{homogen}) can be interpreted  as requiring $f$ and $\Omega$  to be homogeneous functions of degree one with respect to theses "new" integration constants, while the function $h$ and the determinant of the sector $(t,r)$ of the metric, namely $\mbox{det}_{[tr]}:=-g^2(1-f^2)$, are of degree zero, i. e.
\begin{eqnarray}
&&f(r,\alpha x_M,\alpha y)=\alpha f(r,x_M, y),\qquad \Omega(r,\alpha x_M,\alpha y)=\alpha \Omega(r,x_M, y),\nonumber\\
&& h(r,\alpha x_M,\alpha y)=h(r,x_M, y),\qquad \mbox{det}_{[tr]}(r,\alpha x_M,\alpha y)=\mbox{det}_{[tr]}(r, x_M, y),
\label{homoconds}
\end{eqnarray}
for any $\alpha\in\mathbb{R}\setminus\left\{0\right\}$.

%%%%%%%%%%%%%%%%%%%%%%%%%%%%%%%%%%%%%%%%%%%%%%%%%%%%%%%%%%%%%%%%%%%%%%%%%%%%%%%%%%%%
\section{Conditions for the disformal invariance of cohomogeneity$-1$ metrics}
%%%%%%%%%%%%%%%%%%%%%%%%%%%%%%%%%%%%%%%%%%%%%%%%%%%%%%%%%%%%%%%%%%%%%%%%%%%%%%%%%%%%%%5
In the previous section we have seen that the odd-dimensional Myers-Perry metric with equal momenta remains invariant by means of a disformal transformation with a constant degree of disformality and with a scalar field given by (\ref{sf}).
This result is strongly correlated to the cohomogeneity$-1$ character of the Myers-Perry metric. We will go further in this direction by considering a general class of cohomogeneity$-1$ metrics (not necessarily a vacuum metric), and by identifying the conditions which ensure that its disformed transformation remains invariant
(up to some diffeomorphisms and some redefinitions of the constants of integration). In order to achieve this task, we consider a class of cohomogeneity$-1$ metrics in arbitrary, $D$, dimensions parametrized as follows for latter convenience
\begin{eqnarray}
\label{coh1}
ds^2=\left(-1+g_{tt}(r,a_{\alpha})\right)dt^2+g_{rr}(r,a_{\alpha})dr^2+2\sum_{i=1}^{D-2}g_{(it)}(r,a_{\alpha})\,dt\,dx^i&&+\sum_{i\not=
j=1}^{D-2}g_{ij}(r,a_{\alpha})\,dx^i\,dx^j\nonumber\\
&&+\sum_{i=1}^{D-2}
g_{ii}(r,a_{\alpha})(dx^i)^2,
\end{eqnarray}
where the $a_{\alpha}$ are some constants (like the mass, angular momenta, electromagnetic charges, $\cdots$). Note that we do not consider off-diagonal terms of the form $g_{tr}(r,a_{\alpha})dtdr$ in the ansatz (\ref{coh1}), since these can always be eliminated due to the cohomogeneity$-1$ property of the metric.

Following the same steps as before, a solution of $X=g^{\mu\nu}\partial_{\mu}\phi\partial_{\nu}\phi=-m^2$ will be given by
\begin{eqnarray}
\label{sfphi}
\phi(t,r)=-m t-m \int \sqrt{-
g_{rr}(r,a_{\alpha})\Big(1+\frac{\vert\Delta_{n}(r,a_{\alpha})\vert}{\vert\Delta_{n+1}(r,a_{\alpha})\vert}\Big)}\,dr,
\end{eqnarray}
where $\vert\Delta_{n+1}\vert$ and $\vert\Delta_{n}\vert$ are
the determinants of the following reduced metrics
\begin{eqnarray*}
\vert\Delta_{n+1}\vert=\left\vert
\begin{array}{cccc}
-1+g_{tt}(r,a_{\alpha}) & g_{tx_1}(r,a_{\alpha}) & \ldots & g_{tx_n}(r,a_{\alpha})\\
\\
g_{tx_1}(r,a_{\alpha})  & g_{x_1x_1}(r,a_{\alpha})  & \ldots & g_{x_1x_n}(r,a_{\alpha}) \\
\vdots  &\vdots  & \ddots & \vdots\\
 g_{tx_n}(r,a_{\alpha})&
g_{x_1x_n}(r,a_{\alpha}) & \ldots & g_{x_nx_n}(r,a_{\alpha})
\end{array}
\right\vert,
\vert\Delta_{n}\vert=\left\vert
\begin{array}{ccc}
 g_{x_1x_1}(r,a_{\alpha})  & \ldots & g_{x_1x_n}(r,a_{\alpha}) \\
\vdots  & \ddots & \vdots\\
g_{x_1x_n}(r,a_{\alpha}) & \ldots & g_{x_nx_n}(r,a_{\alpha})
\end{array}
\right\vert.
\end{eqnarray*}
With these definitions, it is clear that the determinant of the cohomogeneity$-1$ metric (\ref{coh1}) denoted by  $\mbox{det}(g)$ is given by
\begin{eqnarray}
\mbox{det}(g)(r,a_{\alpha})=\vert\Delta_{n+1}(r,a_{\alpha})\vert\, g_{rr}(r,a_{\alpha}).
\label{detdef}
\end{eqnarray}
%To see this, we can note that
%\begin{eqnarray}
%g^{tt}=\frac{\vert\Delta_{n}(r,a_{\alpha})\vert}{\vert\Delta_{n+1}(r,a_{\alpha})\vert},\qquad
%g^{rr}=\frac{1}{g_{rr}}.
%\end{eqnarray}
Using the scalar field defined by (\ref{sfphi}), the disformed metric,
\begin{eqnarray}
d\bar{s}_{{\tiny\mbox{disf}}}^2=ds^2-P\left(d\phi\right)^2,
\end{eqnarray}
yields, after eliminating the undesired off-diagonal term $dt-dr$ (which is possible since we have a cohomogeneity$-1$ metric),
\begin{eqnarray}
d\bar{s}_{{\tiny\mbox{disf}}}^2=&&\left(-1-Pm^2+g_{tt}(r,a_{\alpha})\right)dt^2+\bar{g}_{rr}(r,a_{\alpha})dr^2+2
\sum_{i=1}^{D-2}g_{(it)}(r,a_{\alpha})\,dt\,dx^i\\
&&+2\sum_{i\not=j=1}^{D-2}g_{ij}(r,a_{\alpha})\,dx^i\,dx^j+\sum_{i=1}^{D-2}
g_{ii}(r,a_{\alpha})(dx^i)^2,
\end{eqnarray}
where, for simplicity we have defined
\begin{eqnarray}
\bar{g}_{rr}(r,a_{\alpha}):=\frac{(1+Pm^2)\,\vert\Delta_{n+1}(r,a_{\alpha})\vert\, g_{rr}(r,a_{\alpha})}{\vert\Delta_{n+1}(r,a_{\alpha})\vert-Pm^2\vert\Delta_{n}(r,a_{\alpha})\vert}=
\frac{(1+Pm^2)\, \mbox{det}(g)(r,a_{\alpha})}{\vert\Delta_{n+1}(r,a_{\alpha})\vert-Pm^2\vert\Delta_{n}(r,a_{\alpha})\vert}.
\label{dr2}
\end{eqnarray}
Hence, although the form of the metric is preserved by the disformed metric, there is a priori no reason for the latter to be diffeomorphic to the original metric (\ref{coh1}), unless, as we will see now, the metric functions satisfy certain conditions. First of all, by noting that the $dt^2-$term can be rewritten as
$$
\left(-1-Pm^2+g_{tt}(r,a_{\alpha})\right)dt^2=\left(-1+\frac{g_{tt}(r,a_{\alpha})}{1+Pm^2}\right)d\bar{t}^2
$$
with $d\bar{t}=dt\,\sqrt{1+Pm^2}$, it is easy to see that the disformal factor $1+Pm^2$ can be absorbed if the metric function $g_{tt}$ satisfies the following homogeneity condition
\begin{eqnarray}
\label{c1}
\frac{g_{tt}(r,a_{\alpha})}{1+Pm^2}=g_{tt}(r,\bar{a}_{\alpha}),
\end{eqnarray}
where $\bar{a}_{\alpha}$ are some redefinitions of the constants $a_{\alpha}$, i.e.
\begin{eqnarray}
\label{c2}
\bar{a}_{\alpha}=f_{\alpha}\left(a_{\alpha}, P, m^2\right).
\end{eqnarray}
Now, because of the time redefinition, we also need to require that the off-diagonal terms on the $t-$row
must satisfy as well a homogeneity condition given by
\begin{eqnarray}
\label{c3}
\frac{g_{it}(r,a_{\alpha})}{\sqrt{1+Pm^2}}=g_{it}(r,\bar{a}_{\alpha}).
\end{eqnarray}
In addition, even if the $g_{ij}-$ terms are not affected by the disformal transformation, nor by the time redefinition, we still have to demand that they remain invariant under the redefinitions (\ref{c2}), that is
\begin{eqnarray}
\label{c4}
g_{ij}(r,x_k,a_{\alpha})=g_{ij}(r,x_k,\bar{a}_{\alpha}).
\end{eqnarray}
Finally, the condition on the $dr^2-$disformed term (\ref{dr2}) that will ensure the full disformed metric to be diffeomorphic to itself reads
\begin{eqnarray}
\label{condd} \frac{\vert\Delta_{n+1}(r,a_{\alpha})\vert\,
g_{rr}(r,a_{\alpha})}{\vert\Delta_{n+1}(r,a_{\alpha})\vert-Pm^2\vert\Delta_{n}(r,a_{\alpha})\vert}=\frac{1}{1+Pm^2}g_{rr}(r,\bar{a}_{\alpha}).
\end{eqnarray}
Nevertheless, it is simple to see that if the conditions (\ref{c1}-\ref{c4}) are fulfilled, we will automatically have
\begin{eqnarray}
\label{condd2}
\vert\Delta_{n+1}(r,a_{\alpha})\vert-Pm^2\vert\Delta_{n}(r,a_{\alpha})\vert=(1+Pm^2)\vert\Delta_{n+1}(r,\bar{a}_{\alpha})\vert,
\end{eqnarray}
and as a direct consequence, the condition (\ref{condd}) will be
achieved if
\begin{eqnarray}
\label{condd3} \vert\Delta_{n+1}(r,a_{\alpha})\vert\,
g_{rr}(r,a_{\alpha})=\vert\Delta_{n+1}(r,\bar{a}_{\alpha})\vert\,
g_{rr}(r,\bar{a}_{\alpha}).
\end{eqnarray}
From the definition (\ref{detdef}), this last equation is equivalent to requiring the determinant of the
metric to remain invariant under the redefinition of the constants of integration (\ref{c2}), i.e.
\begin{eqnarray}
\label{c5}
\mbox{det}(g)(r,a_{\alpha})=\mbox{det}(g)(r,\bar{a}_{\alpha}).
\end{eqnarray}
To summarize, we have shown that, in order for the cohomogeneity$-1$ metric (\ref{coh1}) to remain invariant under a disformal transformation generated by a scalar field with constant kinetic term, and with a constant factor of disformality, $P$, the metric functions have to satisfy the listed conditions (\ref{c1}), (\ref{c3}), (\ref{c4}) and (\ref{c5}). In particular, these conditions ensure that the hair of the scalar field, $m$, and the constant, $P$, can be absorbed into the redefinitions of the constants (\ref{c2}). It is also easy to check that in the case of the odd-dimensional Myers-Perry metric with equal momenta these conditions reduce to the homogeneous conditions listed previously (\ref{homoconds}).

%%%%%%%%%%%%%%%%%%%%%%%
\section{Conclusions}
%%%%%%%%%%%%%%%%%%%%%%%

The main objective of the present work is to look for non-trivial rotating black hole solutions of some general extended scalar tensor theories. Here, we restrict our study to a general shift symmetric and parity preserving scalar tensor action that contains up to second order covariant derivatives of the scalar field. In order to tackle the problem of finding non-trivial rotating configurations of
the field equations, we fix the metric spacetime to be a vacuum rotating black hole spacetime, and we investigate whether this metric can accommodate a non-trivial scalar field. Since the metric is fixed and corresponds to a vacuum metric, it is reasonable to identify these solutions with stealth configurations.
In order to ensure that the solution will not require any fine-tuning of the
coupling functions, our ansatz for the scalar field is such that its kinetic term is assumed to be constant.
In doing so, we prove that such a scalar field has to satisfy a single
tensorial equation (\ref{cond}), and this solution would exist for any coupling functions of the theory. Our hypothesis on the scalar field is also useful to identify it with the Hamilton-Jacobi potential, and to take advantage on the known results on the integrability of the Hamilton-Jacobi equations.

In four dimensions, it was known that the condition (\ref{cond}) cannot be satisfied in the Kerr  metric. Starting from this observation, we consider the
problem of finding rotating stealth black hole solutions defined on the five-dimensional Myers-Perry metric.
In doing so, we have noticed that such an ansatz is not appropriate, unless we impose some restrictions on the metric, and on the scalar field. In particular, the angular momentum parameters of the Myers-Perry metric must be chosen to be equal. This restriction is known to enhance the symmetry group of the metric to a $U(2)$ symmetry, and allows to express the metric components entirely as functions of a single (radial) coordinate. In addition, we have shown that the scalar field solution must depend linearly on time and on the radial coordinate (\ref{sf}). The non-trivial scalar field solution corresponds to the Hamilton-Jacobi potential in which the energy must be given by the particle mass, the conserved quantities associated to each rotation
must be taken to be zero, and where the equivalent of the Carter's constant is taken to be zero, $K=0$.

Interestingly enough, we have also shown that the disformed cohomogeneity$-1$ Myers-Perry spacetime obtained using this stealth scalar field is diffeomorphic to itself. This means that the hair of the scalar field identified with the particle mass and
the constant disformality parameter can be consistently absorbed into further redefinitions of the mass and of the single angular parameter of the disformed metric. In other words, the invariance of the disformal transformation can be viewed as a map that brings a rotating black hole configuration with mass $M$ and angular momentum $a$ to another rotating configuration with rescaled mass and angular momentum, and where this rescaling is quantified by the hair and the disformality parameter. The invariance of the cohomogeneity$-1$ Myers-Perry spacetime under a disformal
transformation can also be explained from the transformation of the Ricci tensor (\ref{riccibar2}) together with the fact that the condition (\ref{cond}) rescales with an overall factor of $\frac{1}{(1-PX)^2}$ under a disformal a transformation, i.e.
\begin{eqnarray}
\bar{\nabla}_{\mu}\bar{\nabla}_{\nu}\phi\,\bar{\Box}\phi+ \bar{\nabla}^{\lambda}\phi\left(\bar{\nabla}_{\lambda}
\bar{\nabla}_{\mu}\bar{\nabla}_{\nu}\phi\right)=\frac{1}{(1-PX)^2}\Bigg({\nabla}_{\mu}{\nabla}_{\nu}\phi\,{\Box}\phi+ {\nabla}^{\lambda}\phi\left({\nabla}_{\lambda}
{\nabla}_{\mu}{\nabla}_{\nu}\phi\right)\Bigg).
\end{eqnarray}
This would imply that for any vacuum metric with a scalar field satisfying the tensorial equations (\ref{cond}), its disformal transformation
generated by the scalar field would as well be a solution of the field equations (\ref{feqs}). Also, since the kinetic term, $X$, of the non-trivial stealth scalar field is constant, one could easily consider more general disformal and conformal transformations that respect the symmetry $\phi\to\phi+\mbox{cst}$, i.e.
$$
\bar{g}_{\mu\nu}=A(X)\,g_{\mu\nu}-P(X)\,\phi_{\mu}\phi_{\nu},
$$
and this will not affect our results. All these results are shown to hold in higher-odd dimensions, where the Myers-Perry metric with equal momenta is known to be of cohomogeneity$-1$ class. Starting from this observation, we have listed the
conditions on a general class of cohomogeneity$-1$ metrics, ensuring its invariance (up to diffeomorphisms) under a disformal transformation with a constant degree of disformality and with a scalar field with constant kinetic term. Finally, in the appendix, we consider the extension to the five-dimensional Kerr-de Sitter metric, where it is shown that rotating stealth solutions exist, provided some fine tuning of the coupling functions of the extended scalar tensor theory.

%%%%%%%%%%%%%%%%%%%%%%%%%%%%%%
\section*{Acknowledgements}
%%%%%%%%%%%%%%%%%%%%%%%%%%%%%%
 We would like to thank Timothy Anson, Eloy Ay\'on-Beato, Eugeny Babichev and Christos Charmousis for very useful discussions.
OB is funded by the PhD scholarship of the University of Talca. The work of MH has been partially supported by FONDECYT grant 1210889.

%%%%%%%%%%%%%%%%%%%%%%%%%%%%%%%%%%%%%%%%%%%%%%%%%%%%%%%%%%%%%%%%%%%%%%%%
\section*{Appendix : Rotating stealth solution with an Einstein space}
%%%%%%%%%%%%%%%%%%%%%%%%%%%%%%%%%%%%%%%%%%%%%%%%%%%%%%%%%%%%%%%%%%%%%%%%

The rotating stealth black hole solutions defined on the odd-dimensional cohomogeneity-$1$ Myers Perry metric can also be extended in the presence of a cosmological constant with some subtleties as we shall see. In this appendix, we present in detail the five-dimensional case, but its extension to higher odd-dimensions is straightforward.

In order to achieve this task, we consider the following scalar tensor theory
\begin{eqnarray}
S[g,\phi]=\int
d^5x\sqrt{-g}&\Big[K(X)+G(X)R+A_1(X)\left[\phi_{\mu\nu}\phi^{\mu\nu}-(\Box\phi)^2\right]+A_3(X)\Box\phi\,\phi^{\mu}\phi_{\mu\nu}\phi^{\nu}
\nonumber\\
&+A_4(X)\phi^{\mu}\phi_{\mu\nu}\phi^{\nu\rho}\phi_{\rho}+A_5(X)\left(\phi^{\mu}\phi_{\mu\nu}\phi^{\nu}\right)^2\Big],
\label{5actionSST}
\end{eqnarray}
whose field equations for a constant kinetic scalar field reduce to
{\small
\begin{eqnarray}
&&-\frac{1}{2}K(X) g_{\mu\nu}+ K'(X)\phi_{\mu}\phi_{\nu}+G(X)G_{\mu\nu}+G^{\prime}(X) R\phi_{\mu}\phi_{\nu}-\frac{1}{2}A_3(X)\Big[(\Box\phi)^2-(\phi_{\alpha\beta})(\phi^{\alpha\beta})-R_{\alpha\beta}\phi^{\alpha}\phi^{\beta}\Big]\phi_{\mu}\phi_{\nu}\nonumber\\
&&+A_1(X)\Bigg[-R_{\nu\lambda}\phi_{\mu}\phi^{\lambda}-R_{\mu\lambda}\phi_{\nu}\phi^{\lambda}-
\frac{1}{2}g_{\mu\nu}\left[(\Box\phi)^2-(\phi_{\alpha\beta})(\phi^{\alpha\beta})\right]+g_{\mu\nu}\left[R_{\lambda\rho}\phi^{\lambda}\phi^{\rho}\right]
+\phi_{\mu\nu}\Box\phi+ \phi^{\lambda}\phi_{\lambda\mu\nu}\Bigg]\nonumber\\
&&-A'_{1}(X)\left[(\Box\phi)^2-(\phi_{\alpha\beta})(\phi^{\alpha\beta})\right]\phi_{\mu}\phi_{\nu}=0,, ~~
\label{5feqs}
\end{eqnarray}}
and the conserved scalar field current, $\nabla_{\mu}J^{\mu}=0$, with
\begin{eqnarray}
J^\mu &=& 2 \left( G'(X) R - \left[ A_1'(X) + \frac{1}{2} A_3(X) \right] \Big[(\Box\phi)^2-(\phi_{\alpha\beta})(\phi^{\alpha\beta})\Big] + \frac{1}{2}A_3(X)\Big[R_{\alpha\beta}\phi^{\alpha}\phi^{\beta}\Big] + K'(X) \right)\phi^\mu \nonumber \\
 &&- 2 A_1(X) R^{\mu\nu}\phi_\nu.
\label{5Jeqs}
\end{eqnarray}
In analogy with the Myers-Perry case, we consider the
cohomogeneity$-1$ five-dimensional Kerr-de Sitter metric \cite{Hawking:1998kw} satisfying $R_{\mu\nu}=4\lambda g_{\mu\nu}$,
\begin{eqnarray}
ds^2=&&-\frac{\Delta}{\rho^2}\Bigg[dt-\frac{a}{\Xi_a}\left(\sin^2\theta d\varphi+\cos^2\theta d\psi\right)\Bigg]^2+
\frac{\Xi_a\sin^2\theta}{\rho^2}\Bigg[a dt-\frac{\rho^2}{\Xi_a}d\varphi\Bigg]^2+\frac{\Xi_a\cos^2\theta}{\rho^2}\Bigg[a dt-\frac{\rho^2}{\Xi_a}d\psi\Bigg]^2\nonumber\\
&&+\frac{(1-r^2\lambda)}{r^2\rho^2}\Bigg[a^2 dt-\frac{a\rho^2}{\Xi_a}\left(\sin^2\theta d\varphi+\cos^2\theta d\psi\right)\Bigg]^2 + \frac{\rho^2}{\Delta}dr^2 + \frac{\rho^2}{\Xi_a}d\theta^2,
\label{kerrads}
\end{eqnarray}
where we have defined
\begin{eqnarray*}
\Delta=\frac{1}{r^2}(r^2+a^2)^2(1-r^2\lambda)-2M,\qquad \Xi_a=1+a^2\lambda,\qquad \rho^2=r^2+a^2.
\end{eqnarray*}
For an Einstein metric satisfying $R_{\mu\nu}=4\lambda g_{\mu\nu}$, it is easy to see that the field equations (\ref{5feqs}) reduce to
{\small
\begin{eqnarray}
&&\left[-\frac{1}{2}K(X)-6\lambda G(X)+4\lambda A_1(X)X-\frac{1}{2}A_1(X)\Big((\Box\phi)^2-(\phi_{\alpha\beta})(\phi^{\alpha\beta})\Big)\right]g_{\mu\nu}\nonumber\\
&&+\Bigg[K'(X)+20 \lambda G'(X)-8\lambda A_1(X)-A_1'(X)\Big((\Box\phi)^2-(\phi_{\alpha\beta})(\phi^{\alpha\beta})\Big)-\frac{1}{2}A_3(X)\Big((\Box\phi)^2-(\phi_{\alpha\beta})(\phi^{\alpha\beta})-4\lambda X\Big)\Bigg]\phi_{\mu}\phi_{\nu}\nonumber\\
&&+A_1(X)\left(\phi_{\mu\nu}\,\Box\phi+ \phi^{\lambda}\phi_{\lambda\mu\nu}\right)=0.
\label{adscase}
\end{eqnarray}}
We do not pretend to solve these equations in full generality but instead opt for a strategy similar to the asymptotically flat case. Indeed, we will consider a scalar field whose kinetic term is a constant, $X=-m^2$, that is
\begin{eqnarray}
\phi(t,r)=-mt-m\int \frac{r(r^2+a^2)\sqrt{\lambda (r^2+a^2)^2+2M}}{(r^2+a^2)^2(1-\lambda r^2)-2Mr^2}dr\Longrightarrow \phi_{\mu}\phi^{\mu}=-m^2.
\label{sf5d}
\end{eqnarray}
Since we are considering the de Sitter case $\lambda>0$, the scalar field is well defined. Nevertheless, one can see that in contrast with the asymptotically flat case, the scalar field, as defined by (\ref{sf5d}), does not satisfy the tensorial conditions (\ref{cond}), but instead
\begin{eqnarray}
\phi_{\mu\nu}\,\Box\phi+ \phi^{\lambda}\phi_{\lambda\mu\nu}=4\lambda \phi_{\mu}\phi_{\nu}+4m^2\lambda g_{\mu\nu}.
\label{condLambda}
\end{eqnarray}
This might seem like an obstruction, but given the structure of the equations (\ref{adscase}), these can be recast using the relation (\ref{condLambda}) into
\begin{eqnarray}
\left[-\frac{1}{2}K-6\lambda G-6 m^2\lambda A_1\right]g_{\mu\nu}+\Bigg[K'+20 \lambda G'-4\lambda A_1-12 m^2\lambda A_1'-8 m^2\lambda A_3\Bigg]\phi_{\mu}\phi_{\nu}=0,
\label{adscase2}
\end{eqnarray}
where we have explicitly used that $X=-m^2$, as well as the trace of Eq. (\ref{condLambda}) which yields
$(\Box\phi)^2-(\phi_{\alpha\beta})(\phi^{\alpha\beta})=12 m^2\lambda$. Now, since $g_{tr}=0$ while $\phi_{t}\phi_{r}\not=0$, each bracket of (\ref{adscase2}) must vanish independently which in turn implies the conditions
\begin{eqnarray}
{{K(X)=-12\lambda\Big(G(X)-X A_1(X)\Big),\qquad G'(X) + A_1(X) + 3 X A_1' + X A_3(X)=0.}}
\label{qaz}
\end{eqnarray}
In \cite{Takahashi:2020hso} different conditions, for example requiring that $A_1$ and $A_2$ vanish at the constant value of $X$, have been implemented to look for solutions of more general quadratic theories.

It is also straightforward to see that under these restrictions, the current $J^{\mu}$  as defined in (\ref{5Jeqs}) vanishes identically, and hence the equations of motion for the scalar field are well verified. Hence, we conclude that the Kerr-de Sitter metric (\ref{kerrads}) together with the scalar field (\ref{sf5d})
will be a solution of the field equations (\ref{5feqs}-\ref{5Jeqs}) provided that the coupling functions are tied as (\ref{qaz}). 

Finally, as in the asymptotically flat case, the disformed metric generated by the scalar field (\ref{sf5d}) with a constant degree of disformality, $d\bar{s}^2=ds^2-P(d \phi)^2$,  is as well an Einstein metric. Indeed, combining the
equations (\ref{riccibar}) and (\ref{condLambda}), one gets
$$
\bar{R}_{\mu\nu}={R}_{\mu\nu}-\frac{P}{1-PX}\left[4\lambda \phi_{\mu}\phi_{\nu}+4m^2\lambda g_{\mu\nu}\right]=4\lambda g_{\mu\nu}
-\frac{P}{1+Pm^2}\left[4\lambda \phi_{\mu}\phi_{\nu}+4m^2\lambda g_{\mu\nu}\right]=\frac{4\lambda}{1+Pm^2}\bar{g}_{\mu\nu}.
$$
Generalizations to higher odd dimensions can be done with the Kerr-de Sitter metric with equal momenta given in \cite{Chen:2006xh}.

%%%%%%%%%%%%%%%%%%%%%%%%%%%%

\end{document}